# Non-Reciprocity Compensation Combined with Turbo Codes for Secret Key Generation in Vehicular Ad Hoc Social IoT Networks

Gregory Epiphaniou, *MIEEE* Petros Karadimas, *MIEEE* Dhouha Kbaier Ben Ismail, Haider Al-Khateeb, Ali Dehghantanha, *SMIEE* Kim-Kwang Raymond Choo, *SMIEE*

**Abstract**—The physical attributes of the dynamic vehicle-to-vehicle (V2V) propagation channel can be utilised for the generation of highly random and symmetric cryptographic keys. However, in physical-layer key agreement, non-reciprocity due to inherent channel noise and hardware impairments can propagate bit disagreements which have to be addressed prior to the symmetric key generation which is inherently important in social IoT networks. This work parametrically models temporal variability attributes such as 3D scattering and scatterers' mobility and for the first time incorporates such features into the key generation process by combining non-reciprocity compensation with turbo codes. Preliminary results indicate a significant improvement in bit mismatch rate (BMR) and key generation rate (KGR) when compared with sample indexing techniques.

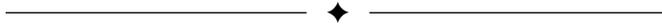

## 1 INTRODUCTION

CONVENTIONAL cryptographic solutions in wireless communications generate shared secrets using pre-computational techniques or asymmetric cryptographic protocols with additional challenges imposed upon energy efficiency, computational complexity and processing-communication overhead during secret key establishment in autonoumos communication of IoT nodes and within social IoT networks [1]. Existing cryptographic solutions are designed independently to the physical properties of the network in which they are applied. This has initiated research activities in the area of fast and efficient key generation algorithms based on physical layer characteristics falling under the broad RSS-based, frequency selectivity-based and CIR-based design approaches [2], [3], [4]. In previously mentioned approaches the wireless channel acts as a medium to increase key generation rate, cryptanalytic resistance, and quality of keys generated between end points due to the inherent stochastic nature of wireless propagation channels [5]. In addition, the ability to generate cryptographic keys using these approaches is removing the necessity to rely on higher-layer encryption protocols. These "channel-based key" extraction approaches try to exploit the physical properties of Wireless channels such as reciprocity and temporal/spatial variability in an attempt to provide the necessary randomness for symmetric key generation [6], [7]. In a typical VANET environment the wireless links between nodes and co-existent adversaries experience un-correlated channel attributes. Therefore, these channels can offer a certain degree of confidentiality during the key generation process between parties. The immediate effect of this, is to reduce computational complexity and relax certain barriers related to key management requirements. The secret key information is usually generated from one or more channel characteristics as part of the signal quantisation phase. The process to determine appropriate channel metrics to characterise a unique wireless channel still remains a challenging and complex domain of scientific enquiry in its infancy [8], [9]. A trade-off also exists between quantisation performance and selection of thresholds with a direct impact (positive or negative) to the key generation rate. The unification of the shared secret key must also adhere to error correction principles and valid processes around privacy enhancement techniques in order to minimise information leakage during message exchanges. This process assures symmetric operation between peers and confidentiality assurance by minimising information exchange for the process of correcting bit mismatch between transceivers.This is especially important in social IoT networks considering autonomous nature of the nodes communicating potentially private information.

In this paper, for the first time we are incorporating all the essential vehicle-to-vehicle (V2V) communication characteristics such as three-dimensional (3D) multipath propagation and surrounding scatterers' mobility (i.e. other vehicles) in the key generation process. Our key generation technique can be used to establish secure communication channels within ad hoc social vehicalar networks. We employ the comprehensive parametric stochastic V2V channel model presented in [10] to synthetically generate the receiver's channel response (Bob's channel), from which

———————————————
G. Epiphaniou, D. Kbaier and H. Al-Khateeb are with the School of Computer Science and Technology, University of Bedfordshire, UK (e-mail: gregory.epiphaniou@beds.ac.uk, Haider.Al-Khateeb@beds.ac.uk, dhouha.kbaier@beds.ac.uk)
P. Karadimas is with the School of Engineering, University of Glasgow, Scotland, (e-mail:Petros.Karadimas@glasgow.ac.uk)
A. Dehghantanha is the School of Computer Science and Engineering, The University of Salford, UK (e-mail: A.Dehghantanha@Salford.ac.uk)
K-K R Choo, is with Department of Information Systems and Cyber Security, The University of Texas at San Antonio,San Antonio, Texas, USA (e-mail: raymond.choo@fulbrightmail.org)

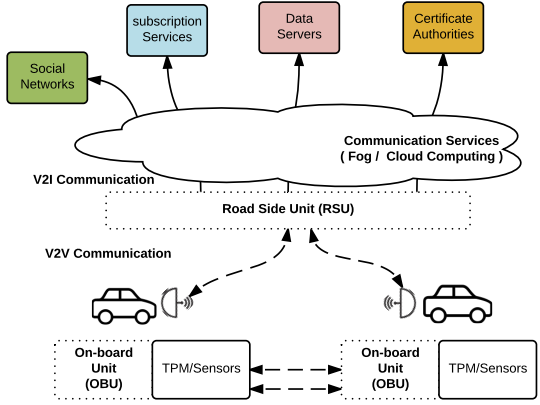

Fig. 1. Vehicular Networking Architecture

the transmitter's response arises after applying the non-reciprocity compensation technique presented in [11]. After the necessary thresholding used to allocate bits according to designated signal levels, we apply for the first time in such setting (V2V channels with parametric 3D multipath propagation and scatterers' mobility) turbo coding (TC) techniques for information reconciliation. Significant improvement in certain key performance indicators (KPIs) has been achieved compared to the existing standard indexing technique described at [12]. For fair comparisons, that indexing technique was again applied in conjunction with the non-reciprocity compensation technique in [11]. More specifically, the key generation rate (KGR) and bit mismatch rate (BMR) are significantly improved when combining both non-reciprocity compensation and TCs in our work.

The rest of this paper is structured as follows: Section 2 reviews existing works in secret key extraction focusing on error reconciliation techniques. In Section 3 we briefly present the performance metrics employed in similar works. In Section 4, we present the adopted key generation process by applying TCs and non-reciprocity compensation in V2V communication channels incorporating 3D multipath propagation and scatterers' mobility. Comparisons with a standard indexing technique are further presented. Finally, Section 5 concludes this paper.

## 2 RELATED WORKS

In vehicular Ad Hoc Networks (VANETs) (See Fig.1), nodes are distributed and self-organised with the majority of wireless communication carried out by on-board units (OBU) integrated with additional services and processes running [13]. High mobility of these nodes and signal diffraction and diffusion properties render these environments susceptible to faster fading, multipath delay, path loss and increased Doppler frequency shift. These unique temporal and spatial properties can generate significant randomness in secret-bit extraction and key distribution while multipath components of the radio channel are identical between two end points at any given time. Also, the prediction of randomness in these dynamic environments is more difficult that static ones due to the high entropy bits extracted in shorter time

[14]. Different approaches have been published in the public domain in secure key extraction protocols with different strengths and limitations with regards to entropy, secret bit extraction rate, key generation rate, number of nodes and threat models. For an exhaustive comparison of these protocols, readers are encouraged to see work in [15].

The secret key information is usually generated from one or more channel characteristics as part of the signal quantisation phase, including fluctuations of signal amplitudes and channel phase [16], [17], [9]. A trade-off exists between quantisation performance and selection of thresholds with a direct impact (positive or negative) to the key generation rate, entropy and bit mismatch rate. These metrics can be affected by the time difference between channel estimates at Alice and Bob, channel decorrelation in time (channel coherence time), inherent communication noise and hardware impairments [18]. The unification of the shared secret key must also adhere to error correction principles and valid processes around privacy enhancement techniques in order to minimise information leakage during message exchanges. Specifically in V2V communications very high temporal variability takes place due to mobility of transmitter, receiver and surrounding scatterers [10], [19], [20]. Though disadvantageous for communication purposes, such temporal variability can be readily exploited in the key generation process. Signal strength variations due to dynamically changing environments have been leveraged in secret key extraction in [21], [22]. Authors have demonstrated certain degree of entropy in the key generation and exchange process under the assumption that an adversary has unbounded capacity to estimate RSS values of the packets transmitted. In [23], authors introduced a filtering technique promised to maintain entropy and improve signal correlation between communication parties by restricting bit generation only for the period of time that that high motion-related fluctuation is present. Movement characteristics and their influence in RSS variation have also been exploited for key generation in [18], [24]. The correlation between the probing rate and key generation rate was observed in [25]. Authors introduced an adaptive probing scheme that dynamically changes the probing rate subject to channel-related parameters. Authors in [26] positively correlate entropy of secret bits as a function of mobility with high secret-bit extraction rate although a single channel observation can lead to lower average number of secret bits generated whereas the authors in [27] model the upper bound of the average secret key extraction rate as a function of the signal bandwidth. Most of the approaches rely on the assumptions that Eve cannot jam the communication channel and is not close to either Alice or Bob.

Additional challenges have been recorded when Receiver's Signal Strength (RSS) is used as a metric to be quantised [11]. Typical thresholds selected usually do not account for points in between them thus reducing the overall key quality or information available for the key generation process. In addition, RSS is usually extracted by a single frequency resulting in low bit generation rates. On the other hand, channel-phased quantisation presents several benefits as higher level of secrecy can be achieved by the uniform distribution of the phases on the channel taps and increase key generation rate by leveraging the whole channel impulse response (CIR) [15]. It is also noticed that a higher number of secret bits can be extracted that removes the need to estimate RSS over a certain time window. RSS-based approaches though do not require significant hardware modifications with better overall performance in respect to synchronisation errors. The CIR can be described as follows [5]

$$h(t) = \sum_{i=0}^{L-1} h_l \delta(t - t_l) \quad (1)$$

where $\delta$ is the impulse function, L is the number of channel paths, $h_l$ is the l-th path complex gain and $t_l$ is the delay of the signal on the l-th path in the multipath channel. The multipath fading channel properties in frequency domain have also been investigated in the literature as an alternative way to achieve high entropy and key generation rate. Channel state information extracted from OFDM subcarriers has been also introduced in an attempt to reduce random noise and improve overall key generation rate [11]. Multiple thresholds are also used to further quantise these average values of channel response to generate a binary sequence. That bit sequence is then normalised through error reconciliation techniques to assure symmetric and identical bits within the key space. Although this approach is generic, applies more on static nodes and does not depend on mobility aspects making it suitable for wireless sensor networks. A further challenge would be the violation of orthogonality due to Doppler effect inherent in VANETs [28].

Authors in [11], argue that channel state information extracted within the coherence time of the channel could be non-reciprocal due to different electrical properties of wireless devices including antenna systems and RF front circuitry. This unavoidably prevents the extraction of symmetric cryptographic keys with low-bit mismatch rate. However, the channel response in different subcarriers should be different due to diversified frequencies. The location and time in which channel response measurements were taken for a specific subcarrier also differ which can be argued as a factor increasing key randomness. Authors in [29] added that channel information at the receiver can be modelled as a location-dependent variable with enough information entropy to be utilised in key generation. However, if channel response is measured in a short period of time highly correlated estimates are generated in both transmitters. A channel gain complement (CGC) algorithm was introduced in an attempt to reduce the disparity of channel responses. The non-reciprocity components were identified with the use of probe packets for each subcarrier. Authors have recorded high bit mismatch rate when channel state information is quantised in the time domain compared to the frequency domain.

The randomness of signal envelope to share the secret key between two parties has also been examined where deep fades have been used to extract correlated bit strings based on a theoretical analysis and simulation results only [30], [18]*. Multiple antenna diversity has also been investigated for secret key extraction with limitations in the key generation rate [31]. Authors* have argued that the signal envelope can provide (to a pair of transceivers) enough entropy required to extract a cryptographic key for data exchange without the necessity to experience identical signal envelops between transceivers. Although focus on deep fades can partially overcome interference problems, however, the quality of the symmetric key and the key generation rate is low. Authors also limit their discussion on the secure ways that key verification information can be exchanged. They also hold assumptions that the size of the bit streams between the two transceivers are the same although calculated by different random sources. Also, work in [30] proved to be computationally expensive when it comes to key recovery phase that render the algorithm difficult to be implemented in V2V communications. Their fuzzy information reconciliation algorithm seems to remove these constraints but the outcome is reduced entropy in the overall quality of the key produced. Information reconciliation is the process of correcting mismatch bits of the quantisation phase by publicly exchanging information to be used for corrective actions [32].

Quantisation and thresholding are the most important processes in the key establishment process as they provide initial information based on channel characteristics. Also, these processes directly affect the bit mismatch probability due to non-fully reciprocal but highly correlated channel responses of Alice and Bob as a result of inherent communication noise and transceivers hardware impairments. The number of thresholds selected during quantisation also presents a tradeoff between key generation rate and random noise. Additional issues with fixed and multiple thresholds were also reported such as susceptibility to active attacks and discard of sampled values between thresholds respectively [5]. Protection against active attacks has been partially addressed in [2] with an Adaptive Secret bit Generation (ASBG) scheme. In this approach sampled values were divided into blocks and each block has been independently quantised using its own thresholds based on its average and standard deviation. Although this work seem to improve overall key generation does not account for imperfect channel reciprocity.

Specifically in V2V communications very high temporal variability takes place due to the mobility of transmitter, receiver and surrounding scatterers. Though disadvantageous for communication purposes, such temporal variability can be readily exploited in the key generation process. Two different techniques have been introduced in [33] namely least square thresholding and neural network-based error reconciliation. Authors recorded an improvement in the detection of fades with smaller depth in environments with no deep fades (e.g., line-of-sight situations). The latter technique uses two similar bit strings to generate keys of arbitrary length known to both Alice and Bob. The security of this system is based on the assumption that Eve cannot adequately reverse the training process of the neural network. A low-cost approach with regards to channel sampling effort was introduced in [25]. The authors modelled mathematically an adaptive channel probing approach based on Lempel-Zin and proportional-Integral-Derivative (PID) controller. Adaptation of the probing rate showed improvements in both the key generation rate and efficiency of the probing process.

The last step in the key generation process assumes that the information extraction about the shared key used should be computationally expensive to adversaries (privacy amplification). Most of the existing approaches focus on different threat models and assumptions around level of access to the channel. "Trapdoor" functions are used as means to assure certain level of authentication and integrity in this process [34]. In the next, we present an overview of the most important error correction codes that can be potentially used in the information reconciliation stage.

## 2.1 Error correction codes

Error reconciliation is the next step in the secret key generation process to correct miss-matched information due to imperfect reciprocity and random noise in the channel. Several error reconciliation algorithms have been introduced with different tradeoffs between communication and computational complexity and throughput error correction capabilities (e.g. Cascade and Winnow). The Cascade error reconciliation protocol assumes that two legitimate parties agree on a random permutation over a public channel [35]. This random permutation takes place over their shifted keys in an attempt to evenly distribute errors. Their shifted keys are then divided in blocks where each block does not present more than one error based on the error rate calculated [36].

Linear error correction codes known as Hamming codes have been also introduced in the literature [37]. In order for a sender to transmit a message with a Hamming code the dot product of a generator matrix and the message must be calculated (code word). The code word is then transmitted at the receiver who computes the product of the code word and the parity check matrix (syndrome). If the calculated syndrome at the receiver is a zero vector, the message was received without any errors. In Winnow protocol [38], the operation is much similar with Cascade. The protocol also suggests privacy maintenance throughout the whole reconciliation phase as a mean to protect information exposed during parity and syndrome exchanges.

Low Density Parity Codes (LDPC) are known for the low density of their parity check matrices which linearly increases the complexity of the decoding algorithm as the length of the message increases [39]. In LDPC codes the minimum distance (as in Hamming codes) and the decoding algorithm used are considered essential parameters to their performance. In their original form LDPC codes have fixed number of 1's in each column k and each row j along with the block n, known as (n,j,k) low density code. The original algorithm developed by Gallager to generate those LDPC matrices was deemed insufficient for large key spaces and limited to work only with regular codes (codes with fixed number of 1's in both columns and rows). LDPC can be more efficient than Cascade as they can become rate adaptive leading to more efficient interactive reconciliation protocols [40], [41].

The invention of turbo codes (TCs) [42] was a revival for the channel coding research community. Historical turbo codes, also sometimes called Parallel Concatenated Convolutional Codes (PCCCs), are based on a parallel concatenation of two Recursive Systematic Convolutional (RSC) codes separated by an interleaver. They are called "turbo" in reference to the analogy of their decoding principle with the turbo principle of a turbo compressed engine, which reuses the exhaust gas in order to improve efficiency.

The turbo decoding principle calls for an iterative algorithm involving two component decoders exchanging information in order to improve the error correction performance with the decoding iterations. This iterative decoding principle was soon applied to other concatenations of codes separated by interleavers, such as Serial Concatenated Convolutional Codes (SCCCs) [43], [44], sometimes called serial turbo codes, or concatenation of block codes, also named block turbo codes [45], [46]. The near-capacity performance of turbo codes and their suitability for practical implementation explain their adoption in various communication standards. In [47] the authors proposed utilizing Turbo codes for reconciliation purposes. Further investigation in [48] show that TCs are good candidates for reconciliation. The efficacy of TCs with regards to their error correction capabilities in various wireless communication standards is also recorded in [49]. Further work in [20] demonstrate the improved performance of TCs over Reed Solomon and CCs which are the de-facto error correction codes used in 802.11p vehicular networks. However, this work does not incorporate comprehensively physical propagation characteristics such as 3D scattering and scatterers' mobility which is addressed in our work.

## 3 PERFORMANCE METRICS

As VANETs are inherently rapidly time-varying due to multipath propagation, this work parametrically models and quantifies such temporal variability attributes and incorporates them into the key generation process. In addition, violation of reciprocity due to hardware impairments or other penalty factors will be compensated in the architectural design and implementation. The proposed algorithmic process will have to compensate for penalty factors influencing the coherence region. The necessity for this work stems from the research effort to further reduce bit mismatch rate while maintaining high key generation rate in practical VANET environments where mobility of the nodes and large network scale imposes unique security challenges. Three performance indicators namely, entropy, secret bit extraction rate and bit mismatch rate, are discussed. The later determines the rate at which the V2V channel is probed in order to secure highly uncorrelated successive samples. We thus present in the following the probing rate together with the three performance indicators.

### 3.1 Probing Rate

The probing rate for both Alice and Bob $f_P = f_{PA} = f_{PB}$ are considered the same for the purpose of channel estimates collection. To achieve uncorrelated successive channel probes, thus achieving highest entropy, successive probes have to be taken in different coherence regions. Thus, we must define $f_P \leq u_{max}$, where $u_{max}$ is the maximum Doppler frequency shift [10]. Considering single bounce of multipath power onto mobile scatterers (e.g., other vehicles), it is defined as [10]

$$u_{max} = \frac{f_c}{c}(u_{Tmax} + u_{Rmax} + 2u_{Smax}) \qquad (2)$$

where $f_c$ is the carrier frequency, $c$ the speed of light in free space and $u_{Tmax}$, $u_{Rmax}$ and $u_{Smax}$ the maximum velocities of transmitter, receiver and mobile scatterers, respectively. In order to maximise the bit extraction rate, we should investigate the feasibility of defining $f_P$ as equal to $u_{max}$.

## 3.2 Entropy measures

The de-facto metric which quantifies the uncertainty is the entropy of the generated bit string. The higher the entropy the limited the ability to deduce a secret key established by Eve due to larger uncertainty introduced. Entropy per bit $i$ is defined as [5]

$$H_i = -p_0 log_2 p_0 - (1-p_0) log_2 (1-p_0) \qquad (3)$$

where $p_0$ the probability of having zero and $1 - p_0 = p_1$ the probability of having one. Ideally, we should have $p_0 = p_1 = 0.5$. For independent bit sequences, the total entropy is $H_{total} = \sum_{i=1}^{N} H_i$, where N is the total number of bits in a sequence [50]. In an ideal case, $H_{total} = N$ bits.

## 3.3 Secret bit extraction rate

The rate is measured in terms of the final secret-bits extracted after error reconciliation and privacy amplification. In practice the secret bit extraction rate depends on the probing rate from Alice and Bob and the number of secret bits per probing. The amount of secret bits extracted in a time varying channel is influenced by the thresholding. Considering 0s and 1s to be generated with equal probabilities (after proper thresholding) the secret bit extraction rate will be $R_k$ [12]

$$R_k = 2 f_P p(A=1, B=1) \qquad (4)$$

where $p(A=1, B=1)$ is the joint probability of having 1 simultaneously at Alice's and Bob's bit strings. However, in this paper we consider key generation rate as the number of symmetric keys produced per unit time.

## 3.4 Bit mismatch Rate

Usually BMR will be measured as a ratio of the number of bits that do not match between Alice and Bob to the number of bits extracted at the thresholding stage often used as a performance criterion for the quantisation process [5]. The BMR is measured immediately after the thresholding stage because a single mismatch in the bitstring can render the secret key unusable. Bit mismatch rate differs from the bit error rate in communication theory, which represents the number of bits received in error. The two reasons for bit mismatch are the unavoidable inherent noise in any wireless communication link and the violation of reciprocity due to hardware impairments. As violation of non-reciprocity is compensated we are left with the inherent noise as a unique problem. This noise will add uncertainty to the transmitted bit strings given the received bit strings. Ideally, both bit strings should have been identical. The bit mismatch probability can be described as follows [12]

$$P_N = 1 - (1-p_e)^N \qquad (5)$$

where $p_e$ will be the probability of a single erroneous bit defined as [30]

$$p_e = P(B=0|A=1) = \frac{P(B=0, A=1)}{P(A=1)} \qquad (6)$$

where $P(B=0|A=1)$ is the conditional probability of Bob's bit being 0 when Alice's is 1.

## 4 NON-RECIPROCITY COMPENSATION AND TC RECONCILIATION IN VANET

The key generation process presented in Fig. 2 considers for error reconciliation the method presented in [12] and for a first time TCs in a V2V environment. However, the input data in our case are generated synthetically in order to comply with V2V propagation settings.

### 4.1 V2V channel model

The synthetic simulated Bob's channel response is generated by employing the Monte Carlo simulation method [51]. For the V2V setting the theoretical channel model that needs to be simulated has been described in detail in [10]. Thus Bob's response in time domain is written as

$$G_B(t) = \sum_{l=1}^{L} |\alpha_l| exp(j\phi_l) exp(j2\pi v_l t) \qquad (7)$$

The Doppler frequency $v_l$ is determined by

$$v_l = v_{T,l} + v_{S,l} + v_{R,l} \qquad (8)$$

where $v_{T,l}$, $v_{S,l}$ and $v_{R,l}$ are the contributions due to Tx mobility, scatterers' mobility and Rx mobility, respectively. The Doppler shift $v_{T(R),l}$ results from the departure (arrival) of the $l^{th}$ multipath component from the mobile Tx (to the mobile Rx). It is defined as [10]

$$v_{T(R),l} = v_{T(R)max} cos\beta_{T(R),l} cos\alpha_{T(R),l} \qquad (9)$$

where $v_{T(R)max} = u_{T(R)}/\lambda$, $\lambda$ is the carrier wavelength, $u_{T(R)}$ the the Tx (Rx) velocity, $\alpha_{T(R),l}$ the azimuth angle of departure (AOD) (angle of arrival (AOA)) and $\beta_{T(R),l}$ the elevation AOD (AOA) with respect to the Tx (Rx) motion. $\alpha_{T(R),l}$ counts from the value $-\pi$ in the negative Y axis returning to the same point in the clockwise direction and $\beta_{T(R),l}$ is zero on the X-Y plane, $\pi/2$ on the positive Z axis and $-\pi/2$ on the negative Z axis. Considering interaction of the $l^{th}$ multipath component with a single mobile scatterer, the Doppler shift $v_{S,l}$ will be [10]

$$v_{S,l} = (u_{S,l}/\lambda)(cos\alpha_{l,l} + cos\alpha_{2,l}) \qquad (10)$$

where $u_{S,l}$ is the scatterer's velocity, $\alpha_{l,l}$ the AOA and $\alpha_{2,l}$ the AOD with respect to scatterer's motion.

The target is to appropriately model each factor affecting the V2V channel response namely $\{|\alpha_l|\}$, $\{v_l\}$, $\{\phi_l\}$. In this paper we consider a normalised (power equal to unity) Rayleigh V2V channel with partially uniform 3D scattering at both Alice's and Bob's sides with a Weibull distribution of the mobile scatterers' velocity. Rather than just a scenario for demonstration, the partially 3D uniform scattering can be further generalized to represent any multipath propagation scenario [52] whereas the Weibull distribution for the multipath power contributed by mobile scatterers has been proved a suitable modeling approach [53]. Thus the scatterers velocity, which in fact models the power contributed by mobile scatterers, is defined as

$$p_{u_s} = w u_S^{b-1} exp(-w u_S^b / b) \quad (11)$$

where $b \leq 1$ is the shape parameter and $w$ the scale parameter. The amplitudes $|\alpha_l|$ are constant and phases $\phi_i$ are uniformly distributed in $[-\pi, \pi]$, i.e., $|\alpha_l| = \sqrt{2/L}$ and $\phi_l \sim U[-\pi, \pi]$ [51]. Each Doppler contribution of Eq. 7 has the following parameters need to be modelled: azimuth angle of departure (AOD), angle of arrival (AOA) $\alpha_{T(R),l} \sim U[A_{T(R)min}, A_{T(R)max}]$ elevation AOD (AOA) $\beta_{T(R),l} \sim U[B_{T(R)min}, B_{T(R)max}]$, AOA to mobile scatterer $\alpha_{1,l} \sim U[-\pi, \pi]$, AOD to mobile scatterer $\alpha_{2,l} \sim U[-\pi, \pi]$, power contributed by mobile scatterers $u_S \sim p_{u_s}(u_S)$. The symbolism $U[.,.]$ stands for the uniform distribution in the designated interval. This scenario can approximate an urban environment with other mobile vehicles and heavy scattering.

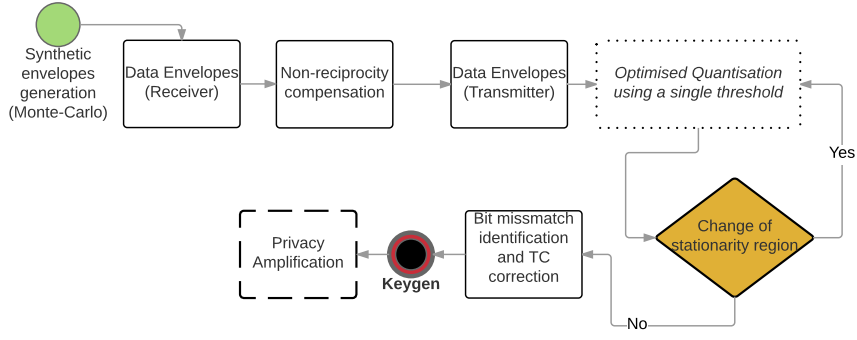

Fig. 2. Algorithmic process for combined TC and NR compensation

The number of multipath components in Eq. 7 is L=20, the sampling/probing rate $F_p = 1/T_{cmin}$ where $T_{cmin} = 1/v_{max} = \lambda/(u_{Tmax} + u_{Rmax} + 2u_{Smax})$ is the minimum coherence in time and $u_{Tmax}, u_{Rmax}, u_{Smax}$ are the maximum Doppler shifts due to mobile transmitter, receiver, and scatterers respectively. In this way, we secure that the channel is mostly probed in different coherence regions, thus successive bits will be independent, resulting keys with maximum entropy. We can further reduce $F_P$ as $1/T_{cmin}$ is in fact its upper bound, however doing so, will reduce the key generation rate, resulting marginal improvement in the key entropy. The latter is just our perception and further research is required, however it goes beyond the scope of this article, which focuses on the applicability of TCs at the information reconciliation stage and potential performance improvement. A possible solution might be to adapt $F_P = 1/T_{cmin}$ to fit in changes of the coherence region due to variations in the propagation conditions (e.g., more intense scatterers' mobility, more directional propagation, etc).

### 4.2 Algorithmic Process

Alice's channel response would normally arise by similar channel probing rate in time instances such that hers and Bob's responses are taken within the same coherence region. However, to further improve performance, Alice's response $G_A(t)$ will arise after applying the non-reciprocity compensation model presented in [11]. Thus considering $M$ estimates within the same coherence region between Alice and Bob, their channel responses are related as [11]

$$G_A(t) - G_B(t) \sim N(0, \sigma^2) \quad (12)$$

The variance is estimated by the discrepancy of Alice's and Bob's estimates as follows

$$\sigma^2 = \frac{1}{M} \sum_{i=1}^{M} (G_{A,i}(t) - G_{B,i}(t) - \mu_t)^2 \quad (13)$$

where

$$\mu_t = \frac{1}{M} \sum_{i=1}^{M} (G_{A,i}(t) - G_{B,i}(t)) \quad (14)$$

This method was presented in [12] where Alice and Bob determine samples from channel estimates above and below an upper and lower threshold discarding those in between, i.e., lossy thresholding. We use this approach to compare it against our TC correction process presented in Figure 2. Those estimates are samples in a form of an excursion. The quantisation process creates segments of those samples (also referred as excursions) of succesive bit values of 1s and 0s. Each of those segments are created whenever a channel probe returns a reading that does not fall inside the thresholds. Alice selects a random set of these segments and sends to Bob the index of the channel estimate lying in the center of the segment defined as $i_{center} = \lfloor \frac{i_{start} + i_{end}}{2} \rfloor$ as a list $L_a$. The number of channel estimates are modelled in the simulation and the total size for each segment has been setup to $m = 5$ successive estimates that fall outside the thresholds (acceptable estimates). However, $m$ is a configurable parameter of the algorithm. For each index from Alice, Bob checks his segments and verifies his samples centered around that index above or below the thresholds $q-, q+$ matched with Alice and generates a new list of those indices $L_b \leq L_a$. Bob sends $L_b$ over to Alice. Both Alice and Bob quantise their channel estimates at each index of $L_b$ in order to generate the bit-string. Thus, this method simultaneously accomplishes thresholding and information reconciliation.

### 4.3 Results and discussion

Part of the algorithmic operation is to develop an optimisation sub-routine to adaptively change the threshold as a function of the temporal variability of the channel. The optimisation routine will consider several attributes such as multi-clustered three dimensional scattering, specular-reflected multipath components, multiple bounces on mobile objects in dense propagation environments. Threshold selection has to be adopted dynamically to the temporal variations induced by the aforementioned effects. The thresholds should be refreshed after a specific amount of time over which the stationarity region has been crossed. We anticipate the refresh to take place after 10 coherence regions because the stationarity region varies. An alternative way to refresh the thresholding process could be to consider a Doppler spectrum correlation criterion. More specifically, considering the normalised Doppler spectrum as a probability distribution of Doppler frequencies, the Doppler correlation coefficient will be defined as

$$\rho(X,Y) = \frac{cov(X,Y)}{\sigma_X \sigma_Y} \quad (15)$$

where $cov(X,Y)$ is the covariance of the X,Y normalised Doppler spectra and $\sigma_X, \sigma_Y$ are the standard deviations of X, Y, respectively. When the correlation coefficient falls below a specified threshold (e.g.,) the quantisation and thresholding process will be refreshed. The first phase of the routine developed is the construction of the Synthetic data which will be generated via Monte Carlo simulation taking into account the number of multiple components, the sampling rate and total number of samples. In the next stage the probed received envelopes are generated considering an appropriately defined probing rate in order to maximize the entropy in the subsequent quantisation step. From the received data, the transmitted data are modelled by considering non-reciprocity compensation. At this stage a lossy quantisation process is preferred due to its computational simplicity. The target is to end up with a maximum secret bit extraction rate and entropy. For that purpose, in the following step several runs should take place considering the thresholds multiple pairs. A feasibility study of both lossless and lossy quantisation processes and their applicability in V-V scenarios is an area for further investigation. Bob's generated sequence after quantization is fed to the input of a TC. During this process a single threshold is adopted as a lossless quantisation scheme with the potential to substantially increase the key generation rate [30]. Turbo decoding is then performed in order to generate a symmetric output, i.e. symmetric keys for Alice and Bob. Performance of the reconciliation method can be evaluated by measuring the BMR and to the Bit Error Rate (BER) in our case. The comparison is made against the sample indexing technique already applied in our algorithm as discussed in subsection 4.2. We calculated BMR for the indexing method by considering the discarded indexes after Alice's and Bob's channel probing. In Table 1 we compute the key generation rate for different key lengths. Compared to the samples' indexing method in [9], there was a significant improvement on both BMR and key generation rate. The simulated BER to generate a symmetric shared key between Alice and Bob after error reconciliation is estimated to only 0.0752 using TCs. Furthermore, the BMR with single thresholding is only 0.02 whereas the estimated BMR with the indexing technique is around 0.22 in both cases of static and mobile scatterers. The key generation rate was also reported high considering different key lengths requested. For instance, the secret key rate to generate the 128-bit symmetric key is 35 good keys per minute with TCs while it varies from 3 to 7 symmetric keys per minute with the indexing technique. As shown in Table 1, simulations proved similar improvements for different key lengths as part of the error reconciliation process. Satisfactory entropy values were obtained throughout all rounds of simulation during the key extraction process ranging from $0,85 \sim 0,97$ bits per sample. Note that the BMR with the indexing technique is nearly the same for different key lengths which is coherent with the uniform method used by the authors. In Table 2, we present a comparison between the BMR achieved in our approach with existing RSS-based approaches published in the literature.

TABLE 1
TC simulation results in secret key generation

| Key Length (bits) | KGR (with TCs) | KGR (with Indexing [12]) |
|---|---|---|
| 128 | 35 keys/min | 3 to 7 keys/min |
| 256 | 17 keys/min | 2 to 5 keys/min |
| 512 | 8 keys/min | 1 to 2 keys/min |

TABLE 2
Comparison of BMR with existing RSS-Based approaches

| Scheme | Design Approach | BMR |
|---|---|---|
| Patwari et al. [54] | RSS-based | 0.482 |
| Jana et al. [14] | | $0 \sim 0.55$ |
| Premnath et al. [2] | | $0.02 \sim 0.24$ |
| Croft et al. [55] | | $0.01 \sim 0.07$ |
| Zan et al. [3] | | $0.005 \sim 0.02$ |
| Mathur et al. [12] | | 0.22 |
| Non-reciprocity compensation with TC (Our approach) | | 0.02 |

## 5 CONCLUSION

We successfully combined non-reciprocity compensation and TCs for information reconciliation as the most important features in V2V communication including 3D scattering and scatterers' mobility. Results have shown significant improvements in key generation rate with reduced BMR when TCs are employed against an existing indexing method. Our proposed technique can be used in for secure communication between vehicular nodes in an ad hoc social IoT network. In future studies, we would like to further investigate TCs for error conciliation purposes especially in the context of social IoT networks. We will focus on several parameters that affect performance of TCs such as component decoding algorithms, number of decoding iterations, generator polynomials, constraint lengths of the component encoders and the interleaver type. Increasing the number of iterations in the TC can significantly improve the BER, thus generating more symmetric keys. Furthermore, we are working towards the single thresholding process by creating a dynamic threshold that is updated according to the receiver's samples.

## ACKNOWLEDGMENT

This work was partially funded by the Defence Science and Technology Laboratory (DSTL), under contract CDE 41130. Moreover, this work is partially supported by the European Council 268 International Incoming Fellowship (FP7- PEOPLE-2013-IIF) grant. The authors would also like to thank Mr George Samartzidis for his initial contribution in the algorithm development.

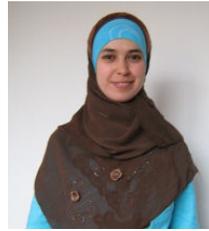

**Dr Dhouha Kbaier** joined the University of Bedfordshire as a Lecturer in Telecommunications and Network Engineering in March 2016. She received from Telecom Bretagne (Brest, France) both her PhD in 2011 with the highest honours and her Master of Engineering degree in 2008. She was specialised in Space Communications Systems in the French Grande cole ISAE in Toulouse, heart of the European Aerospace Industry. Prior to working at the University of Bedfordshire, she worked for several years as a post-doctoral research follower first at Telecom Bretagne, then with Thales Airborne Systems and finally at IFREMER.Thanks to her multi-disciplinarily and her diverse research background, Dr Kbaier was awarded in February 2016 by two French lecturer qualifications in two different fields. She is Fellow of the Higher Education Academy and an Engineering Professors Council (EPC) member. Her research was particularly awarded by several productivity bonuses and an IEEE best paper award. Her research interests include signal processing applied to telecommunications and oceanography, channel coding, digital communications and information theory, error correction in VANET environments, etc.

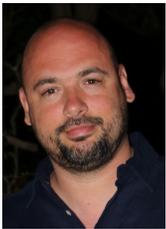

**Dr Gregory Epiphaniou** has been a leading trainer and developer for bespoke Cyber Security programmes with a dedicated, strong team of experts and trainers in several technical domains in both offensive and defensive security. He has also contributed to a numerous public events and seminars around cyber security, course development and effective training both private and government bodies. He currently holds a position as a senior lecturer in Cybersecurity at the University of Bedfordshire. He also holds several industry certifications around Information Security, and currently acts as a subject matter expert in the Chartered Institute for Securities and Investments.

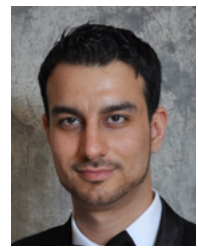

**Dr Haider Al-Khateeb** specialises in Cyber Security, Digital Forensics and Incident Response (DFIR). He holds a first-class BSc (Hons) in Computer Science and PhD in Cyber Security. He is a university lecturer, researcher, consultant, trainer and a Fellow of the Higher Education Academy (FHEA), UK. Haider has published numerous professional and peer-reviewed articles on topics including authentication methods, IoT forensics, cyberstalking, anonymity and steganography. He is a lecturer in the School of Computer Science and Technology and conducts research within the Institute for Research in Applicable Computing (IRAC), University of Bedfordshire.

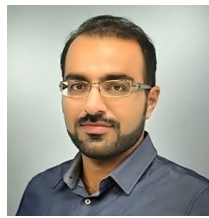

**Dr. Ali Dehghantanha** is a Marie-Curie International Incoming Fellow in Cyber Forensics, a fellow of the UK Higher Education Academy (HEA) and an IEEE Sr. member. He has served for many years in a variety of research and industrial positions. Other than Ph.D in Cyber Security he holds several professional certificates such as GXPN, GREM, GCFA, CISM, and CISSP.

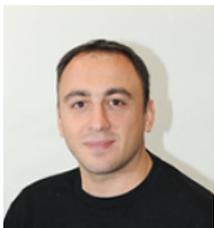

**Dr Petros Karadimas** was born in Greece. He completed his Diploma (MEng) and PhD degrees in the Department of Electrical and Computer Engineering, University of Patras, Greece, in 2002 and 2008, respectively. In December 2009, he was appointed as a Research Fellow in the Centre for Wireless Network Design (CWIND), Department of Computer Science and Technology, University of Bedfordshire, UK, and in October 2011, he was appointed as a Lecturer in Electronic Engineering in the same Department, where he was promoted to Senior Lecturer in August 2015. In August 2016, he moved to the University of Glasgow, UK, as a Lecturer in Electronics and Electrical Engineering. His ongoing research interests and publications are in the fields of Wireless Channel Characterization and Multi-Antenna Systems Performance. Very recently he started researching on Wireless Security over Physical Layer and Wireless Receiver Design Aspects (e.g., OFDM, CDMA systems).

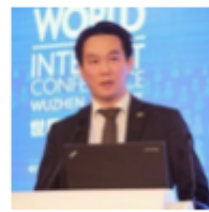

**Kim-Kwang Raymond Choo** received the Ph.D. in Information Security in 2006 from Queensland University of Technology, Australia. He currently holds the Cloud Technology Endowed Professorship at The University of Texas at San Antonio, and is an associate professor at University of South Australia. He is the recipient of various awards including ESORICS 2015 Best Paper Award, Winning Team of the Germany's University of Erlangen-Nuremberg (FAU) Digital Forensics Research Challenge 2015, and 2014 Highly Commended Award by the Australia New Zealand Policing Advisory Agency, Fulbright Scholarship in 2009, 2008 Australia Day Achievement Medallion, and British Computer Society's Wilkes Award in 2008. He is a Fellow of the Australian Computer Society.